\documentclass[nologo,url,11pt,a4paper]{ETHpaper}

\usepackage{fancyhdr,epsfig,url,hyperref,lastpage}
\usepackage[dvips]{color}
\usepackage{amsmath}

\title{Personalised and Dynamic Trust in Social Networks}

\author{Frank E. Walter, Stefano Battiston, and Frank Schweitzer}

\address{Chair of Systems Design, ETH  Zurich, Kreuzplatz 5, 8032 Zurich, Switzerland \\
  \url{fewalter@ethz.ch}, \url{sbattiston@ethz.ch}, \url{fschweitzer@ethz.ch}}

\reference{Submitted to Recommender Systems 2009 (May 09, 2009).}

\www{\url{http://www.sg.ethz.ch}}

\begin{document}

\maketitle

\begin{abstract}
  We propose a novel trust metric for social networks which is
  suitable for application to recommender systems. It is personalised
  and dynamic, and allows to compute the indirect trust between two
  agents which are not neighbours based on the direct trust between
  agents that are neighbours. In analogy to some personalised versions
  of PageRank, this metric makes use of the concept of feedback
  centrality and overcomes some of the limitations of other trust
  metrics.  In particular, it does not neglect cycles and other
  patterns characterising social networks, as some other algorithms
  do.  In order to apply the metric to recommender systems, we propose
  a way to make trust dynamic over time. We show by means of
  analytical approximations and computer simulations that the metric
  has the desired properties. Finally, we carry out an empirical
  validation on a dataset crawled from an Internet community and
  compare the performance of a recommender system using our metric to
  one using collaborative filtering.
\end{abstract}

Personalisation, Information Overload

\section{Introduction}
\label{sec:introduction}

An increasing number of information technologies focuses on how web
users can effectively share opinions about various types of products,
services or even other users. These technologies are the basis of
several types of Web 2.0 applications such as collaborative tagging,
social bookmarking \cite{cattuto07, golder05} and, in particular, also
recommender systems. Given the \textit{heterogeneity of web users}, a
major issue is how to appropriately aggregate opinions in order to
provide judgements that are useful for each individual user.

Most of these applications use collaborative filtering algorithms
which compute an index of \textit{similarity} between users or between
items, based on the ratings that users have provided on these items
\cite{goldberg92,herlocker99,montaner03a}.  When a user belongs to a
community with common, shared tastes, these algorithms work well in
suggesting new items similar to the ones the users have already
rated. There are several other benefits: except providing enough
ratings, no further action is required of users; algorithms for
collaborative filtering are scalable (when similarities are computed
across items \cite{sarwar01}); and, finally, they provide some level
of personalisation.  A shortcoming is that if users are looking for
items which are seldomly rated by their community, the predictions are
poor -- e.g. people who have rated only travel books may not receive
very good recommendations on tools for gardening.

To cope with this, a line of research has focused on basing
recommendations for users not on their similarity, but on their
\textit{trust relations} to other users. In this context, trust is
meant to be the ``expectancy of an agent to be able to rely on some other
agent's recommendations'' \cite{marsh94,walter08-jaamas}. There has been a
body of work on ``trust webs''
\cite{abdul-rahman00,grandison00,marsh94,sabater05} and on their
application to recommender systems
\cite{golbeck05,massa06,montaner02b}. The small-world property of
social networks \cite{newman02} allows to potentially reach a lot of
information, while the trust allows to filter out the relevant pieces
\cite{walter08-jaamas}. The benefits of these trust-based algorithms
include strong personalisation, no need to have a long rating history
in the system because recommendations are not based on similarity, and
the ability to receive recommendations on items different from the
ones already rated.  Some limitations of the trust-based approach
concern the scalability and the fact that, in addition to their
ratings of items, users have to provide information about their level
of trust to some other users.

In this paper, we introduce a novel metric for trust in social
networks. A trust metric allows to compute the indirect trust between
two agents in a social network which are not neighbours, based on the
direct trust between agents that are neighbours. While it is intuitive
to do this on a chain, e.g.~from user $A$ via user $B$ to user $C$,
for instance by multiplying the values of trust along the chain, it is
not a priori trivial how to proceed when a graph contains multiple,
redundant paths, cycles, or triangles (because of mathematical issues
related to uniqueness and consistency). This is a crucial issue
because these patterns all play an important role in social networks,
in particular for the diffusion of information and the build-up of
social capital \cite{wassermann94,vega-redondo07}. Some trust metrics
address these issues by reducing the direct trust graph to an acyclic
graph before applying their computation of indirect trust
\cite{golbeck05,massa06}. Other metrics use only the path of the
shortest distance or of the highest trust \cite{walter08-jaamas}.  Our
trust metric takes all the paths in the graph into account and it is
well-defined on any given graph. It provides each user with
personalised trust ratings about other users in the network. Our
metric also is dynamic, i.e.~it evolves in time depending on how
useful the information received by users is to them. This makes the
metric suitable for application in recommender systems, as we will
illustrate in the remainder of the paper.

\section{Background and Motivation}
\label{sec:background-motivation}

Consider a scenario in which there is a social network of agents which
have trust relationships among each other. This can be described by a
graph in which the nodes represent the agents and the links represent
the trust relationships. There also is a set of objects which can be
rated by agents. Since each agent only knows a few objects, it may
want to know other agent's opinions on unknown objects. However, since
there are potentially many opinions of other agents, it needs to be
able to determine which of these are trustworthy. This implies that an
agent needs to reason about the trustworthiness of other agents
\cite{walter08-jaamas}. However, since its time and resources are
constrained, an agent can only build and maintain trust relationships
with a limited number of other agents.

\textit{Thus, if $T_{ij} \in [0,\,1]$ represents the level of direct
  trust of agent $i$ towards $j$, how do we compute the indirect trust
  $\tilde{T}_{kl}$ between two agents $k$ and $l$ that are not
  neighbours\footnote{Variables expressing indirect trust are as the
    corresponding ones expressing direct trust, but with a tilde
    symbol: e.g.~$T$ and $\tilde{T}$.}?  }

In the following, we will describe the TrustWebRank metric for
computing indirect trust in a network with direct trust. This metric
builds on the concept of feedback centrality which assigns a
centrality score to the nodes of a network based on the centrality
scores of the node's neighbours. In other words, in feedback
centrality, the higher (or lower) the centrality score of a node's
neighbours, the higher (or lower) this node's own centrality is.
These principles can be adapted to define a metric for the
trustworthiness of agents in a social network with trust
relationships.

We briefly review PageRank, one of the most widely known and studied
feedback centrality algorithms \cite{brin98,brandes05}.
In our scenario this would mean to assign a trustworthiness score
$c_i$ to an agent $i$ that depends on the trustworthiness of its
neighbours $j$ (adapted from \cite{brandes05}):
\begin{eqnarray}
  c_i = \beta \sum_{\{j : i \in N_{j}\}}{\frac{c_j}{|N_{j}|}} + (1-\beta) \quad \forall i,
  \label{eq:definition_page_rank}
\end{eqnarray}
where $N_i$ is the set of neighbours of $i$, and $\beta$ is a damping
factor which is chosen around $0.8$ \cite{brin98}. In vector notation:
\begin{eqnarray}
  c = \beta Pc + (1-\beta)\mathit{1},
  \label{eq:eq:definition_page_rank_matrix}
\end{eqnarray}
where $P$ is a stochastic\footnote{We will always assume
  row-stochastic when we state ``stochastic''; this does not imply
  that the matrix need (or not) to be column-stochastic.
} transition matrix defined as
\begin{eqnarray}
  P_{ij} = \left\{ \begin{array}{ll}
  \frac{1}{|N_j|} & \textrm{if there exists a link from\ } j \textrm{\ to \ } i \\
  0 & \textrm{otherwise.} \\
  \end{array} \right.
  \label{eq:eq:definition_page_rank_transition_matrix}
\end{eqnarray}
Eqs.~(\ref{eq:definition_page_rank}) and
(\ref{eq:eq:definition_page_rank_transition_matrix}) can easily be
extended to weighted graphs \cite{brandes05}. Solving Eq.
(\ref{eq:eq:definition_page_rank_matrix}) for $c$ we obtain:
\begin{eqnarray}
  c & = & (I-\beta P)^{-1}(1-\beta)\mathit{1},
  \label{eq:derivation_page_rank}
\end{eqnarray}
where $I$ is the identity matrix and $\mathit{1}$ is the vector
consisting of ones. Since $P$ is, by construction, stochastic and
thus, by the Perron-Frobenius theorem \cite{seneta06}, the largest
eigenvalue is $\lambda_{\mathrm{PF}}(P)=1$, it follows that
$\lambda_{\mathrm{PF}}(\beta P)=\beta<1$. This ensures the existence
of a unique solution of $c$. Usually, one uses Jacobi iteration to
compute such a solution.

The result of applying this algorithm to a graph is a vector which
gives a score of the trustworthiness $c_i$ for each node $i$ in the
graph. Note that this is a \textit{global} metric, i.e.~there is one
score for each agent. It has been observed in the literature that, for
recommender systems, such metrics are often not appropriate and that
\textit{local} metrics, which are \textit{personalised} for each agent
(``how trustworthy is agent $i$ from the perspective of agent $j$''),
are required \cite{massa06}. EigenTrust, for example, is a
PageRank-inspired, global trust metric \cite{kamvar03}.

\section{A Novel Trust Metric}
\label{sec:novel-trust-metric}

\subsection{From Centrality to Trust}
\label{sec:from-centrality-to-trust}

Proceeding in analogy to PageRank and using the principles of feedback
centrality to construct a personalised metric for trust, one could
define the indirect trust of agent $i$ to $j$ as the indirect trust of
the neighbour agents $k$ of agent $i$ to agent $j$, weighted by the
trust of agent $i$ towards these neighbour agents $k$. Let $T$ be the
trust matrix, where $T_{ij} \in [0,1]$ reflects the \textit{direct}
trust from agent $i$ to agent $j$ ($T_{ij}=0$ if there is no link
between agent $i$ and agent $j$). $S$ is the stochastic matrix
\begin{eqnarray}
  S_{ij} = \frac{T_{ij}}{\sum_{k \in N_{i}}T_{ik}},
  \label{eq:definition_normalized_direct_trust}
\end{eqnarray}
where $N_{i}$ is the set of neighbours of agent $i$. $S$ is a
normalisation of $T$. We define $\tilde{T}_{ij}$ to be the
\textit{indirect} trustworthiness score from $i$ to $j$:
\begin{eqnarray}
  \tilde{T}_{ij} = \sum_{k \in N_i}{S_{ik}\tilde{T}_{kj}} \quad \forall i,j
  \label{eq:definition_mole_tidal_trust}
\end{eqnarray}
This allows us to estimate the trust between any two agents $i$ and
$j$: if there is a link between $i$ and $j$, $T_{ij}$ reflects the
trust between them; if there is no link between $i$ and $j$, $\tilde
T_{ij}$ reflects the trust between them. Notice that this definition
is similar to to the approaches used in \cite{golbeck05,massa06}. In
matrix notation, this is the recursive definition
\begin{eqnarray}
  \tilde{T} = S\tilde{T}
  \label{eq:eq:definition_mole_tidal_trust_matrix}
\end{eqnarray}

Notice that this approach has several limitations:

\textit{1) Uniqueness of the solution}: Let $\tilde{v}_{*j}$ be one
column of $\tilde{T}$, i.e.~the vector that expresses how much agent
$j$ is trusted by other agents. Then, Eq.
(\ref{eq:eq:definition_mole_tidal_trust_matrix}) gives
\begin{eqnarray}
  \tilde{v}_{*j}=S\tilde{v}_{*j} & \forall j.
  \label{eq:eq:definition_mole_tidal_trust_vector}
\end{eqnarray}
If $S$ is acyclic \cite{seneta06} (i.e.~the underlying graph is so),
then there is a unique solution of
Eq.~(\ref{eq:eq:definition_mole_tidal_trust_vector}). If $S$ is not
acyclic, it can be either primitive or non-primitive \cite{horn90}. If
$S$ is primitive (and stochastic), there is a unique solution of
Eq.~(\ref{eq:eq:definition_mole_tidal_trust_vector}), a vector with
all components being identical \cite{seneta06}. This would imply that
all agents $i$ would trust agent $j$ equally, which is obviously not
desirable. If $S$ is not primitive, there are multiple solutions for
Eq.~(\ref{eq:eq:definition_mole_tidal_trust_vector}), which also is
not desirable.

One way of dealing with this could be to make $S$ acyclic, for example
by constructing a tree with a breadth-first search (BFS) from a chosen
node, as for example \cite{golbeck05,massa06} do. The BFS selects one
node as a root, and from there on, explores the neighbours of the
nodes, proceeding in levels $1, 2, 3, \ldots$ and removing links within
a level and links from level $k$ to level $l$ where $l<k$ at each
step. However, this entails further limitations:

Social networks are characterised by a high clustering coefficient
\cite{wassermann94,newman02,vega-redondo07}. By making the underlying
graph of a social network acyclic, one removes the links within each
level and the links from levels $k$ to $l$ where $l \leq k$, thus
making the clustering coefficient $0$. This implies that, subsequent
to this procedure, the trust metric will not be able to differentiate
well between regions of high clustering (thus, possibly high trust)
and regions with lower clustering (thus, possibly lower trust) as on
the original graph.

Further, depending on which node is chosen as the root of the BFS, the
acyclic graph will be different. This is not a problem in a
decentralised scenario, when the computation is spread over many
nodes. In this case, each node computes its own set of
$\tilde{v}_{*j}$ by being root of its own breadth-first
exploration. However, this is a problem in a centralised scenario,
where such an approach is not scalable and also not mathematically
tractable: as a result of a BFS rooting at each $i$, the computation
uses a different matrix $T$ for each node.

\textit{2) Combination of direct and indirect trust}: The metric
defined in Eq.~(\ref{eq:definition_mole_tidal_trust}) is not able to
account properly for the following situation: consider an agent $i$
that trusts a neighbour agent $j$ with intermediate level of trust,
e.g.~$T_{ij} \approx 0.5$, because it does not yet know this agent
well. If many of the other neighbours of agent $i$ trust agent $j$,
this should increase the trust between agent $i$ and $j$. This does
not happen with the current definition of trust.

\textit{3) Normalisation of trust}: another property, resulting from
Eq.~(\ref{eq:definition_normalized_direct_trust}), is that the normalisation
removes knowledge from the system. If an agent $i$ trusts $n$
neighbours equally, it does not matter whether it trusts them a lot or
a little in $[0,1]$ -- the normalisation would assign the same value
of trust of $\frac{1}{n}$ to each of the neighbours. Then, during
propagation, only the relative trust compared to other neighbours is
considered. Equally, suppose that an agent $i$ has just one neighbour
agent $j$ -- no matter whether $i$ trusts $j$ highly or lowly, in each
case the normalisation would cause the trust from $i$ to $j$ to be
$1$. The normalisation is necessary, however, to have values of direct
and indirect trust which are in the same range.

\subsection{The TrustWebRank Metric}
\label{sec:trustwebrank}

Thus, given these limitations, can we modify
Eq.~(\ref{eq:definition_mole_tidal_trust}) in such a way that the
following requirements are met?

\textbf{Requirement 1}: The solution of the equation over graphs with
cycles is unique, but not trivial.

\textbf{Requirement 2}: The range of indirect trust is the same as for
direct trust, i.e.~$[0,1]$, so that direct and indirect trust can be
compared.

\textbf{Requirement 3}: In the metric, direct trust ``adds on'' to
indirect trust (capturing the fact that it complements it).

One possibility to address these issues is the following: we compute
the indirect value of trust between two agents $i$ and $j$ based on
the direct trust between them, if there is any, but also based on the
trust that the neighbours of $i$ have in $j$:
\begin{eqnarray}
  \tilde{T}_{ij}=S_{ij}+\beta\sum_{k \in N_i}{S_{ik}\tilde{T}_{kj}} \quad \forall i,j,
  \label{eq:definition_indirect_trust}
\end{eqnarray}
where $\beta \in [0,1)$.
Now, in matrix form Eq.~(\ref{eq:definition_indirect_trust}) is
\begin{eqnarray}
  \tilde{T} & = & S + \beta S \tilde{T},
  \label{eq:eq:definition_indirect_trust_matrix}
\end{eqnarray}
and,  using elementary algebra, we can derive
\begin{eqnarray}
  \tilde{T} & = & (I - \beta S)^{-1}S.
  \label{eq:derivation_indirect_trust}
\end{eqnarray}
There exists a unique, non-trivial solution to
Eq.~(\ref{eq:derivation_indirect_trust}) if
$\lambda_{\mathrm{PF}}(\beta S)<1$, \cite{horn90}. Since $S$ is
stochastic, i.e.~$\lambda_{\mathrm{PF}}(S)=1$, and $\beta \in [0,1)$,
it follows that $\lambda_{\mathrm{PF}}(\beta S)<1$ (Requirement 1).

The parameter $\beta$ has a similar role as the damping factor in
PageRank in Eq.~(\ref{eq:definition_page_rank}): given $\beta \in
[0,1)$, the impact of agents far away in the social network is
discounted. This can be seen more clearly when expressing $(1-\beta
S)^{-1}$ as a geometric sum in
Eq.~(\ref{eq:derivation_indirect_trust}) \cite{horn90}:
\begin{eqnarray}
  \tilde{T}=(1-\beta S)^{-1}S=\sum_{k=0}^{\infty}{(\beta S)^kS}=S+\beta S^2+\beta^2 S^3+\ldots
  \label{eq:role_of_beta}
\end{eqnarray}
The $k$th power of the adjacency matrix of a graph gives the number of
walks of length $k$ between any two nodes in the graph. Similarly, the
$k$th power of the matrix $S$ gives the sum of the products of the
weights along all walks of length $k$ in the underlying graph of
$S$. In Eq.~(\ref{eq:role_of_beta}), the higher the length of the
walks, the stronger the discount (since $\beta<1$). As in PageRank, a
reasonable value of $\beta$ turns out to be around $0.75$ to $0.85$
(see Section \ref{sec:empirical-validation}). Note that $\tilde{T}_{ij}
\notin [0,1]$. We can normalise it to
\begin{eqnarray}
  \tilde{S}_{ij}=\frac{\tilde{T}_{ij}}{\sum_{k \in N_i}{\tilde{T}_{ik}}}
  \label{eq:definition_normalized_indirect_trust}
\end{eqnarray}
to ensure the comparability of values of direct and indirect trust
(Requirement 2).

Furthermore, if agents $i$ and $j$ are not neighbours, the indirect trust
of $i$ to $j$ is entirely based on how much the neighbours of $i$
trust $j$. However, if agent $i$ has a neighbour $j$, the indirect
trust of $i$ to $j$ will also incorporate how much the other
neighbours of agent $i$ trust or do not trust agent $j$ (Requirement
3).

The definition of Eqs.~(\ref{eq:definition_indirect_trust}) and
(\ref{eq:eq:definition_indirect_trust_matrix}) naturally takes the
real structure of a social network into account without needing to
prune any link. Unlike to what would happen during the conversion of
the underlying graph to a tree using a BFS, the algorithm preserves
the links which, in a social network, lead to a high clustering
coefficient, and are not negligible when reasoning about the social
network itself \cite{wassermann94,newman02,vega-redondo07}.

When dealing with huge graphs, however, inverting a matrix as required
by Eq.~(\ref{eq:derivation_indirect_trust}) poses an issue of
computation time and memory. Yet, instead of inverting a matrix or
computing eigenvectors, it is possible to use an iterative method
\cite{brandes05} as follows:
\begin{eqnarray}
  \tilde{T}_{ij}^{(k+1)}=S_{ij} + \beta \sum_{l \in N_i} S_{il} \tilde{T}_{lj}^{(k)} \quad \forall i,j.
  \label{eq:iterative_computation_indirect_trust}
\end{eqnarray}
At each step $k$, one only needs the neighbourhood $N_i$ of a given
agent $i$, as well as access to the matrix of $\tilde{T}^{(k-1)}$
computed at the previous step $k-1$. Notice that now we are computing
a \textit{matrix} while, with the centrality, e.g.~in PageRank, we
were computing a \textit{vector}. This is natural since the centrality
is one value per agent (it is a global notion), while trust is a value
per pair of agents (it is a local, personalised notion). Therefore
computing trust ($\sim O(N^2)$) is inherently more expensive than
computing centrality ($\sim O(N)$). However, do we really need to
compute indirect trust among all agents? In fact, for a given agent
$i$, computing the trust to a selected amount of other agents $j$, if
well chosen, will be sufficient, as the trust to agents far away in
the network will be damped out anyway. So, the scalability of the
trust computation rather is ($\sim O(mN)$), where $m$ is the number of
other agents $j$ to consider for each agent $i$.

\section{An Application of the Metric}
\label{sec:application-of-the-metric}

So far, we have described a trust metric which allows to compute a
measure of trust between two agents which are not necessarily
neighbours in a social network. We will now construct a simple model
which applies this metric in the context of a \textit{recommender
  system}. The purpose is to show how it is possible to compute
predictions of how an agent $i$ likes a particular object $o$ (suppose
a book, CD, or movie) based on how other agents $j$ liked that item
combined with how much $i$ trusts $j$.

\subsection{A Simple Model}
\label{sec:simple-model}

Suppose we have a system of agents embedded in a social network,
defined by a graph and associated to an adjacency matrix $A$. Each
agent $i$ keeps track of its trust relationships to neighbours
$j$. These are reflected in the matrix of direct trust $T$. Obviously,
$T_{ij}>0$ only if $A_{ij}=1$. For the moment, we take the network to
be described by a random graph \cite{erdos59,bollobas85} in which each
agent roughly has the degree~$d$.

Let each agent $i$ be characterised by a profile $\pi_{i}$. The
profile expresses which ratings an agent would give to all possible
objects; however, agents only know a subset of their ratings on
objects. Given an object $o$, $r_i^o \in \{-1,1\}$ is the rating of
agent $i$ on object $o$. If an agent is willing to share all its
opinions with other agents, then the set of all of its ratings
corresponds to its profile; however, there may be agents which are not
willing (because they want to keep their secrets) or able (because
they simply do not know particular objects) to share ratings. This can
be captured by a parameter $\eta$ which reflects the probability of an
agent to share -- i.e.~signal -- its rating with other agents. E.g.,~a
value of $\eta=0.1$ would imply that, on average, at each time step
10\
agents. At the moment, $\eta$ is the same value for all agents, but it
could also be set differently for each agent $i$ or even for each pair
of agents $i$ and
$j$. 

If an agent $i$ is not willing or able to share its rating for an
object $o$, the system computes a prediction $p_i^o$ as follows:
\begin{eqnarray}
  p_i^o=\sum_{j \in N_i}{\tilde{S}_{ij}r_j^o},
  \label{eq:definition_prediction}
\end{eqnarray}
so $p_i^o \in [-1,1]$, since $\sum_{j \in N_i}\tilde{S}_{ij}=1$ and
$r_j^o \in \{-1,1\}$. In vector notation,
\begin{eqnarray}
  p = \tilde{S}r,
  \label{eq:eq:definition_prediction_matrix}
\end{eqnarray}
i.e.~the prediction for an agent $i$ is the sum of the ratings of all
neighbours $j$ weighted by the indirect, normalised trust that agent
$i$ has in these neighbours $j$.

Note that this bears resemblance to Collaborative Filtering (CF)
\cite{goldberg92,herlocker99} in which the prediction for an agent $i$
is also computed as a weighted sum of the ratings of all neighbours
$j$ (not neighbours in a graph-theoretic sense, but neighbours in
terms of similarity of ratings). The more similar a neighbour, the
more influential its rating will be for the prediction. In our case,
making a prediction based on the ratings of the trusted neighbours
implies that we make the assumption that agents who are connected by
trust have similar mind-sets. Notice that this does not imply that
they have rated the same items -- for example, one user could
appreciate the knowledge of another user in gardening, even though his
own domain are travel books. Thus, unlike the similarity that could be
computed e.g.~by Pearson correlation, this notion of similarity
extends not just across rated items, but rather is an ``expected''
similarity reflecting a similar mind-set of two agents.

\subsection{Trust Dynamics}
\label{sec:trust_dynamics}

So far, we have a static model which, based on the trust web of a
particular agent $i$ and the ratings $r_j^o$ of its neighbours $j$, is
able to compute predictions $p_i^o$ for that agent. We now would like
to model the evolution of the trust network over time in the sense
that, based on the quality of a particular recommendation, agent $i$
can update its trust to its neighbours $j$. This adds a time dimension
to the model and requires a mechanism to update the trust between
neighbours. This can be done by adding a utility function: agents
experience a utility by using the ratings or predictions of neighbours
and then the trust update is coupled with the utility experienced. We
define each agent $i$ to experience a utility $u_{ij}(t)$ by following
the recommendation from each neighbour $j$ at time $t$ as follows:
\begin{eqnarray}
  u_{ij}(t) = \left\{ \begin{array}{ll}
  1 - |r_i^o(t)-r_j^o(t)| & \textrm{if\ } j \textrm{\ signals to\ } i \\
  1 - |r_i^o(t)-p_j^o(t)| & \textrm{otherwise}. \\
  \end{array}\right.
  \label{eq:definition_utility}
\end{eqnarray}
Note that $u_{ij}(t) \in [-1,1]$. If the neighbour $j$ signals to
agent $i$, it knows the rating $r_j^o(t)$; otherwise, it only knows a
prediction $p_j^o(t)$.  The closer the recommendation of agent $j$ for
agent $i$ to the rating of agent $i$ is, the greater the agents'
similarity is and thus the higher the utility $u_{ij}(t)$ that agent
$i$ experiences from the recommendation of agent $j$ at step $t$
is. Note that because of the level of cooperation $\eta$ -- which
affects whether agent $j$ signals to $i$ -- the utility takes into
account not only similarity \cite{ziegler06}, but also cooperation
between agents. Based on the utility, agent $i$ can update the trust
towards its neighbour agents $j$. We distinguish four cases, based on
the \textit{sign} and the \textit{magnitude} of the utility:

\begin{itemize}
\item If the sign is positive, this means that the rating or
  prediction of a neighbour was good; if it is negative, it means that
  the rating or prediction was bad.

\item If the magnitude is large, the neighbour had a lot of trust in
  the rating/prediction of its own neighbours; if it is small, the
  neighbour had little trust in the rating/prediction of its own
  neighbours.
\end{itemize}

This leads us to the following definition of how an agent $i$ updates
its trust to agent $j$ from time $t$ to $t+1$:
\begin{eqnarray}
   \breve{T}_{ij}(t+1) = \left\{ \begin{array}{l}
   \gamma T_{ij}(t) + (1-\gamma) |u_{ij}(t)| \\
   \hspace{1ex} \textrm{if\ } u_{ij}(t) > u_{thr} \textrm{\ or\ } -u_{thr} \leq u_{ij}(t) \leq 0 \\
   \gamma T_{ij}(t) - (1-\gamma) |u_{ij}(t)| \\
   \hspace{1ex} \textrm{if\ } u_{ij}(t) < -u_{thr} \textrm{\ or\ } 0 < u_{ij}(t) \leq u_{thr}
   \end{array}\right.
   \label{eq:trust_update}
\end{eqnarray}
where we take $u_{thr}=0.5$ and $\gamma \in [0,1]$ is a parameter that
controls the relative weights of the current history of trust between
two agents, $T_{ij}(t)$, and of the current utility, $u_{ij}(t)$. For
$\gamma>0.5$, this gives the history of trust more weight than the
current utility. In the analysis and simulations (next section), we
found that $\gamma=0.75$ is a reasonable value. Since $\breve{T}_{ij}
\in [-1,1]$, but we want $T_{ij} \in [0,1]$, we cap it to $[0,1]$:
\begin{eqnarray}
   T_{ij}(t+1)=\max(0,\min(1,\breve{T}_{ij})).
   \label{eq:trust_update_minmax}
\end{eqnarray}

\begin{figure*}[t]
  \centering
  \includegraphics[width=0.23\textwidth]{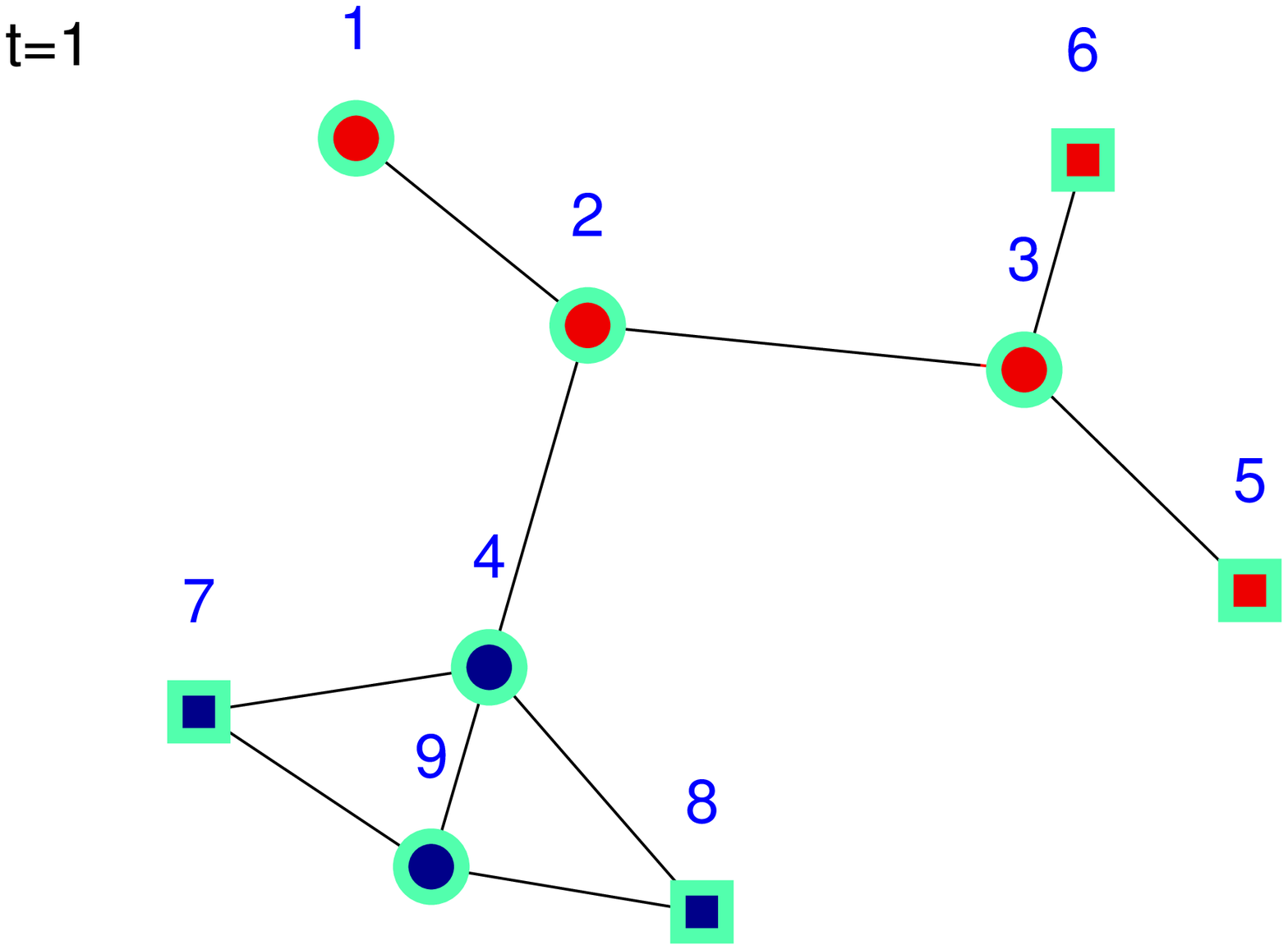}
  \includegraphics[width=0.23\textwidth]{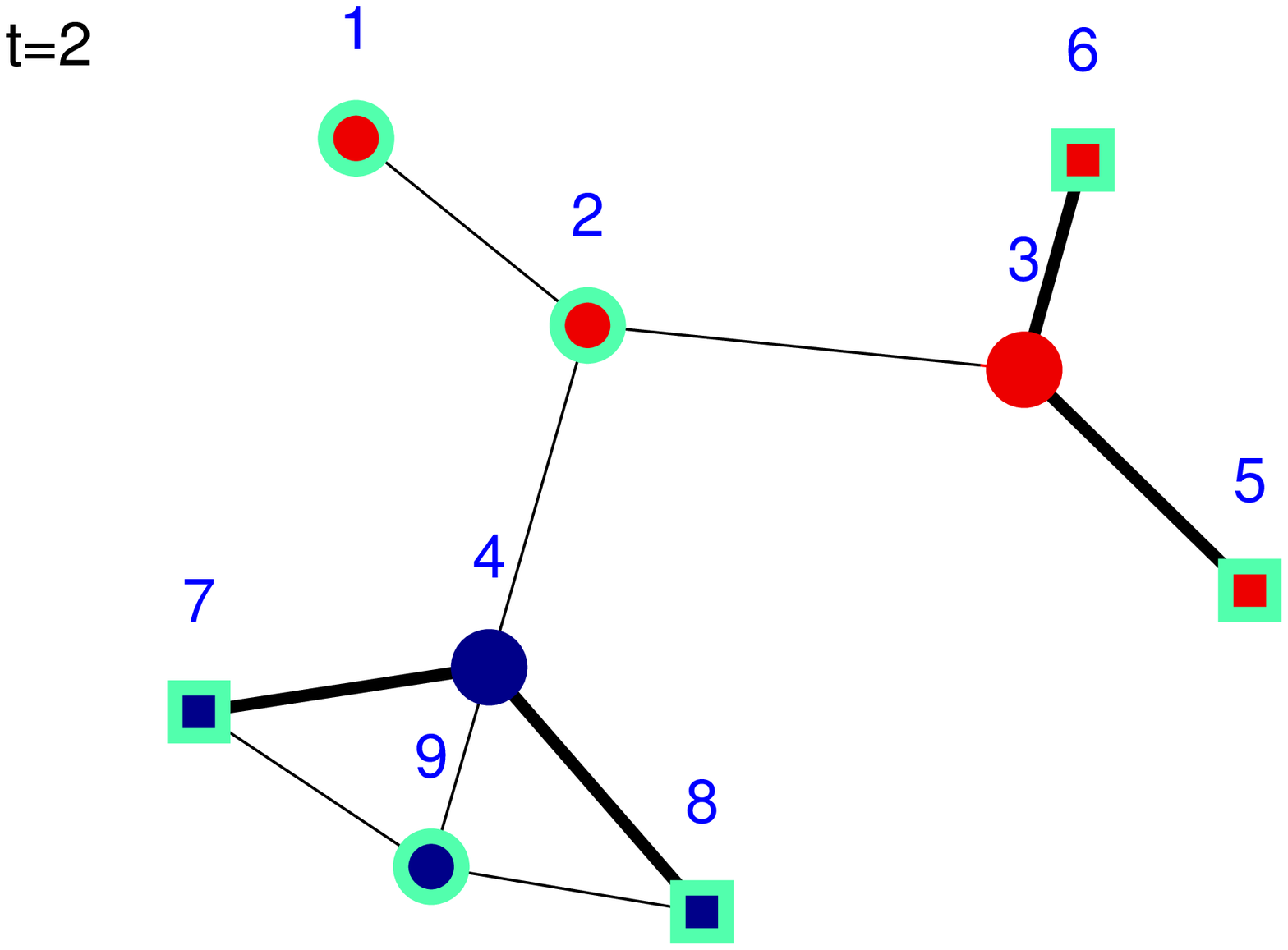}
  \includegraphics[width=0.23\textwidth]{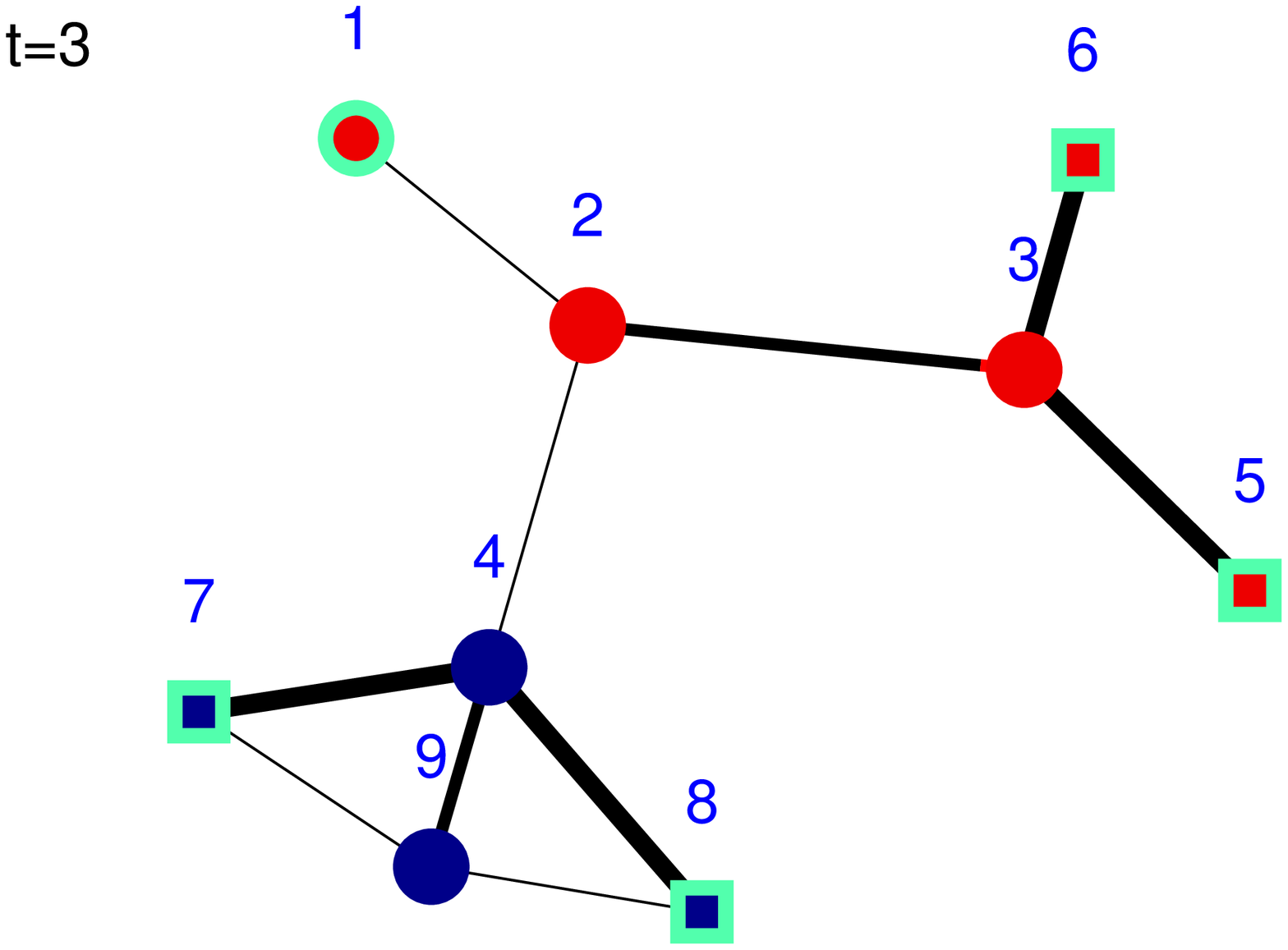}
  \includegraphics[width=0.23\textwidth]{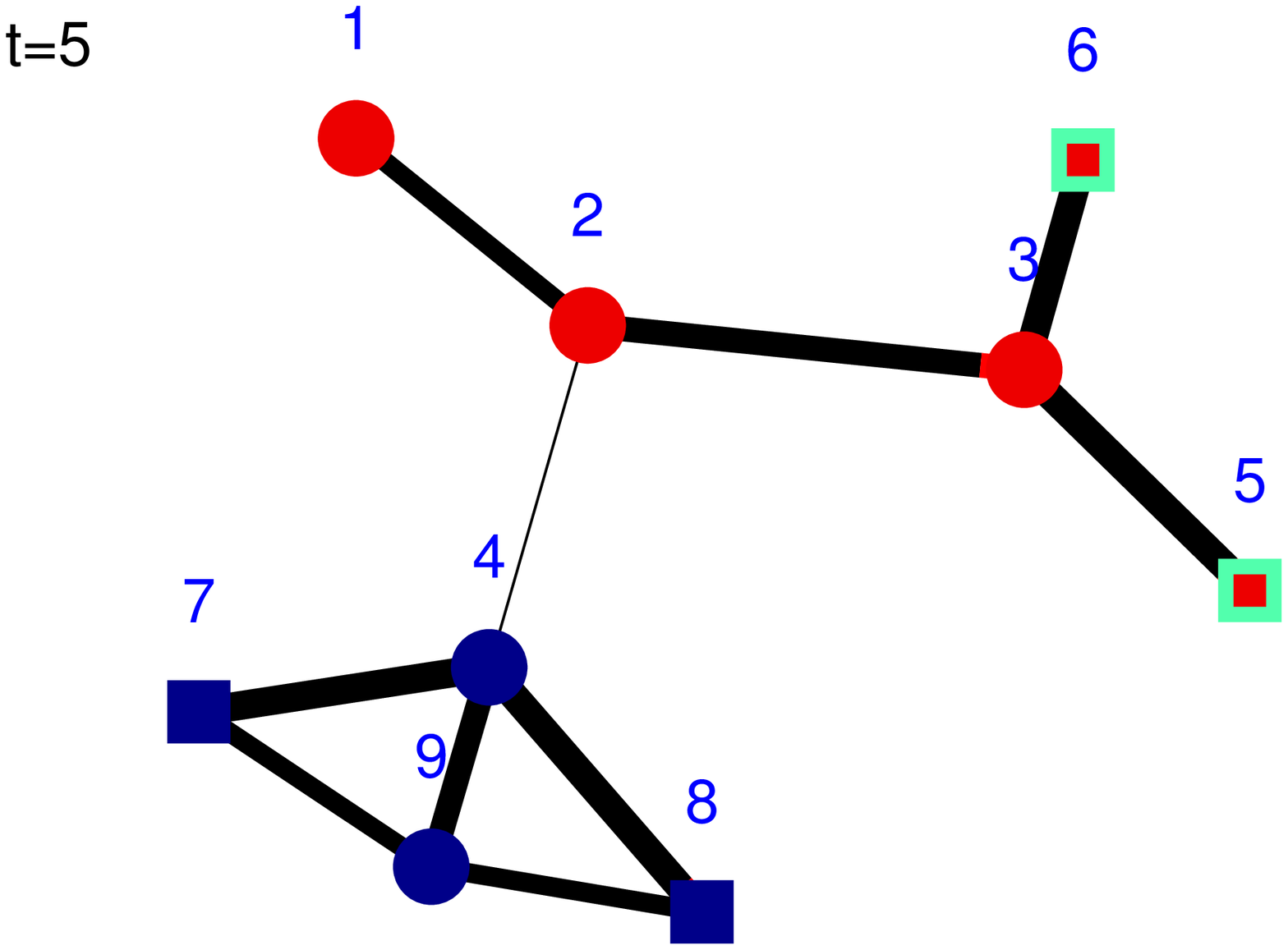}
  \caption{Social network of agents and trust build-up over time in
    case of a fraction of agents not signalling as well as cycles in
    the underlying network: there are two profiles, red and blue,
    indicated by the cores of the node. Only square nodes are
    signalling; e.g., nodes 2, 3, and 4 are not signalling. There are
    two cycles from 4 to 7 and 8, respectively, to 9 and then to
    4. After a few steps, the nodes learn which other nodes to trust.}
  \label{fig:model-illustration}
\end{figure*}

As an example, the effects of these dynamics are illustrated in Figure
\ref{fig:model-illustration}: this is an example of a network of
agents having two profiles (red and blue). Some nodes are signalling
(squares), others are not (circles). The network contains cycles. At
$t=1$, the agents are just connected, the trust between all agents is
equal to zero. At $t=2$, agent 3 and agent 4 have received
recommendations from agents 5 and 6, and from agents 7 and 8,
respectively. Since agent 3 (4) has the same profile as agents 5 and 6
(7 and 8), namely red (blue), it perceives a high positive utility
from the recommendation and thus increases its trust to the
recommending agents. At $t=3$, the system can now provide a
recommendation to agent 2, even though agents 3 and 4 are not
signalling their own rating. Since agent 2 has the same profile as
agent 3, trust between these two agents increases. Agent 2 perceives a
high negative utility from the recommendation of agent 4, thus its
trust remains zero. At the same time, the links from 3 to 5 and 6
reinforce. The same happens in the cycles.  These mechanisms continue
and we see that at $t=5$, paths of trust have developed between agents
of the same profile. Although agent 1 has no agents of its profile
that are signalling in one or two levels of distance, it is still able
to discover a path to two agents of its profile that are signaling and
further away in the network.

\subsection{Analysis of the Model}
\label{sec:analysis-model}

In this section we derive a self-consistent equation for the matrix of
trust which allows to investigate the dynamics of trust. We analyse
the case of a population of agents with only two opposite profiles
(see Section~\ref{sec:simple-model}) which provide ratings on objects
as $+1$ or $-1$, respectively.

We want to compute the expected value of trust at the equilibrium of
the dynamics defined in Eqs.~(\ref{eq:trust_update})
and~(\ref{eq:trust_update_minmax}). We do so by a mean-field
approximation in which we replace the utility $u_{ij}(t)$ in
Eq.~(\ref{eq:trust_update}) with the expected utility over time,
denoted by $u_{ij}:=E(u_{ij}(t))$ (without the time dependency). We
impose $T_{ij}(t)=T_{ij}(t+1)$ at the equilibrium, obtaining
\begin{eqnarray}
  T_{ij}=\max(\min(u_{ij},1),0),
  \label{eq:equilibrium_T_1}
\end{eqnarray}
which requires us to estimate $u_{ij}$. Given the definition of
$u_{ij}(t)$ in Eq.~(\ref{eq:definition_utility}) and the fact that
agents signal a rating with probability $\eta$ and they do not with
probability $1-\eta$, it follows that the expected utility $u_{ij}$ is
\begin{eqnarray}
  u_{ij}= \eta (1 - |\pi_i-\pi_j| ) +(1-\eta)( 1 - |\pi_i-\sum_{k} \tilde S_{jk} \pi_{k}|).
  \label{eq:expected_utility}
\end{eqnarray}
Since we are considering the simple case in which agents signal
faithfully, the expected rating provided by an agent $j$ coincides
with its profile: $E(r_j^o) = \pi_j$. We can thus express the expected
prediction for agent $j$ as $E(p_j^o) = \sum_{k} \tilde S_{jk}
\pi_{k}$. In future work, we will also consider more complicated
cases, e.g.~including non-faithful (selfish or malicious)
behaviour. Substituting into Eq.~(\ref{eq:equilibrium_T_1}), we get:
\begin{multline}
  T_{ij} = \max(0,\min(1,\eta (1 - |\pi_i-\pi_j| ) \\
  + (1-\eta)( 1 - |\pi_i-\sum_{k} \tilde S_{jk} \pi_{k}|))).
  \label{eq:equilibrium_T_3-tmp}
\end{multline}
Since the profiles $\pi$ are given, $T$ is a function of $\tilde
S$. Notice that by combining
Eqs.~(\ref{eq:role_of_beta}-\ref{eq:definition_normalized_indirect_trust}-\ref{eq:definition_normalized_direct_trust}),
we can express $\tilde S_{jk}$ in terms of the components $T_{jk}$,
$(T^2)_{jk}$, $(T^3)_{jk}, \ldots$ as well as $T_{jl}$, $(T^2)_{jl}$,
$(T^3)_{jl}, \ldots$ where $l$ are the other neighbours of $j$:
\begin{eqnarray}
  \tilde S_{jk} 
  &=& \frac{T_{jk} + \beta (T^2)_{jk} + \beta^2 (T^3)_{jk} + \dots}{\sum_{l} \sum^\infty_{m=0}  (T^m)_{jl}}.
  \label{eq:tildeS(T)}
 \end{eqnarray}
 It follows that we can express the value of trust $T_{ij}$ between
 any pair of agents in terms of the value of trust among the other
 pairs. This leads to a self-consistent equation for $T$, where the
 only parameters are the initial values of trust $T(0)$, the
 probability to signal, $\eta$, the discount factor along the walks of
 the graphs, $\beta$, and the profiles of the agents, $\pi$:
\begin{eqnarray}
  T_{ij}= f(T,T(0),\eta,\beta,\pi)\ \forall i,j.
  \label{eq:equilibrium_T_3}
\end{eqnarray}
Notice that Eq.~(\ref{eq:equilibrium_T_3}) is obtained without any
assumption on the structure of the network that is reflected in $T$.

One is, of course, interested in the fixed points of Eq.
(\ref{eq:equilibrium_T_3}), their stability and whether they are
attained by the dynamics. On the one hand, it is trivial to check that
the matrix $T$ with $T_{ij}=1$ among agents with the same profile and
$T_{ij}=0$ among agents with opposite profile is a fixed point of
Eq.~(\ref{eq:equilibrium_T_3}). Denote this configuration as
$\{T^+=1,T^-=0\}$. On the other hand, the configuration with trust
equal zero among all pairs $\{T^{+,-}=0\}$ is not a fixed point.

In the next section, we find, by means of computer simulations, that
the system, starting from a configuration with no trust among the
agents, $\{T^{+,-}=0\}$, always evolves to a configuration in which
agents with similar profile trust each other $\{T^+=1,T^-=0\}$. This
is true even if agents do not signal all the time (i.e. $\eta < 1$).
A formal investigation of the stability of all the fixed points of
Eq.~(\ref{eq:equilibrium_T_3}) will be performed in future work.

\subsection{Simulations}
\label{sec:simulations}

The simulations that we carried out were done on an agent population
of $500$ agents. We considered two opposite profiles with ratings on
objects as $+1$ or $-1$. The agents are connected in a random graph
\cite{erdos59,bollobas85}. Initially, $T_{ij}=0 \quad \forall i,j$,
i.e.~the agents have to learn who to trust. We varied the average
degree $d$ of each agent, as well as the level of cooperation $\eta$
in the system. The following figures illustrate the system behaviour
over $50$ steps; all results were averaged over $100$ runs.

\begin{figure}[h]
  \centering
  \includegraphics[width=0.4\textwidth]{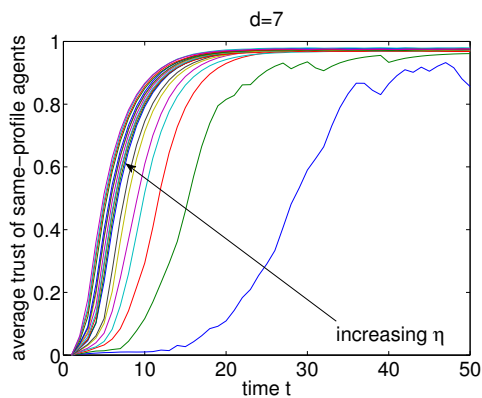}
  \vspace{-1.5ex}
  \caption{Trust between agents of the same profile over time, for a
    fixed average degree of agents but variable level of cooperation.}
  \label{fig:trust-d-n-time}
\end{figure}

Figure~\ref{fig:trust-d-n-time} illustrates the average trust between
agents of the same profile over time: the average degree of agents is
fixed, $d=7$, and the level of cooperation $\eta$ is variable, ranging
from $0.01$ to $0.25$ in steps of $0.01$. The average trust between
agents of the same profile converges to $1$ for almost all $\eta$. For
larger $\eta$, this process takes place much faster than for smaller
$\eta$. Given a sufficient level of cooperation in the system, the
agents develop trust to the agents that have the same
profile. Furthermore (this is not shown in the figure), agents of
opposite profiles do not develop trust between each other.

\begin{figure}[h]
  \centering
  \includegraphics[width=0.4\textwidth]{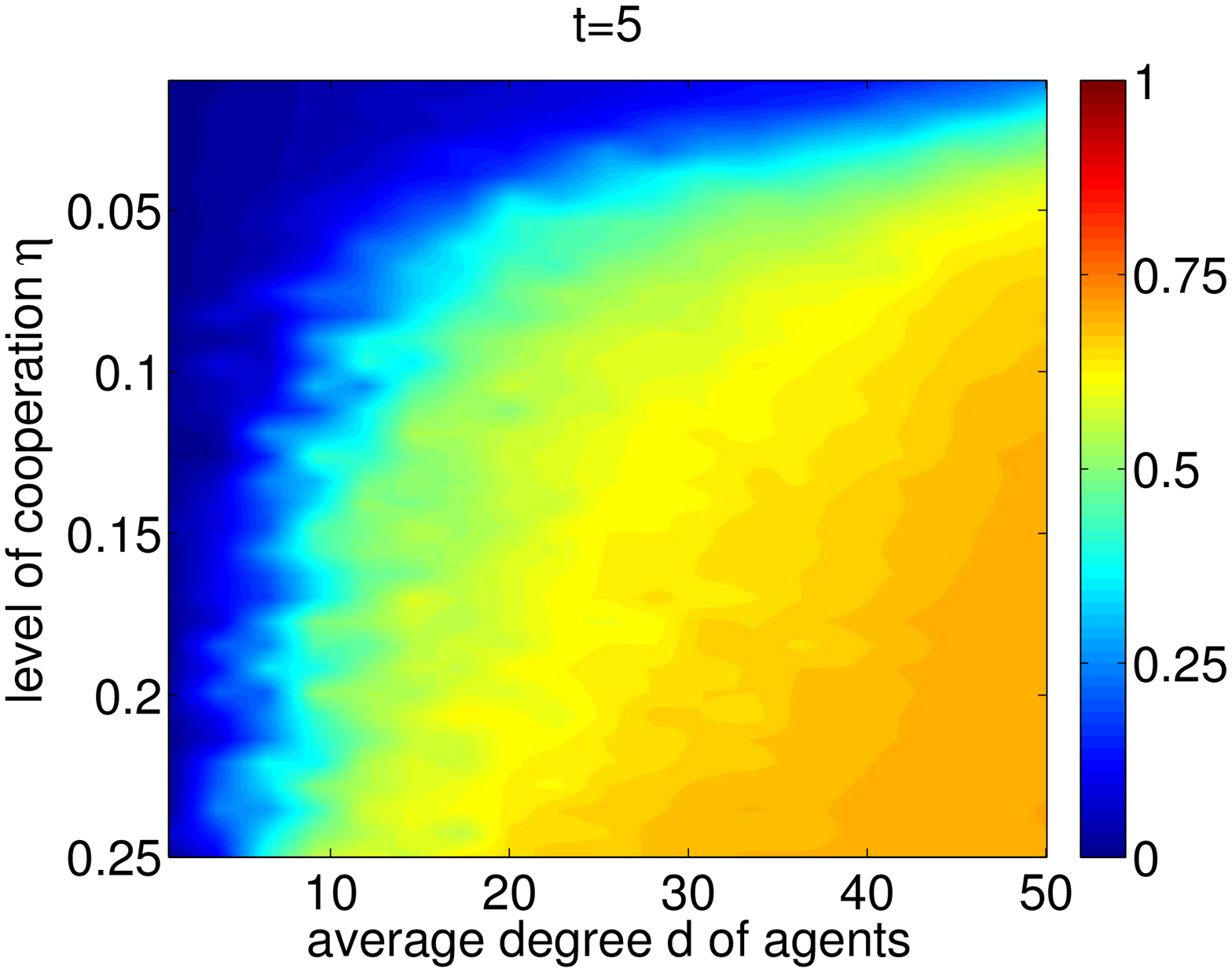}
  \includegraphics[width=0.4\textwidth]{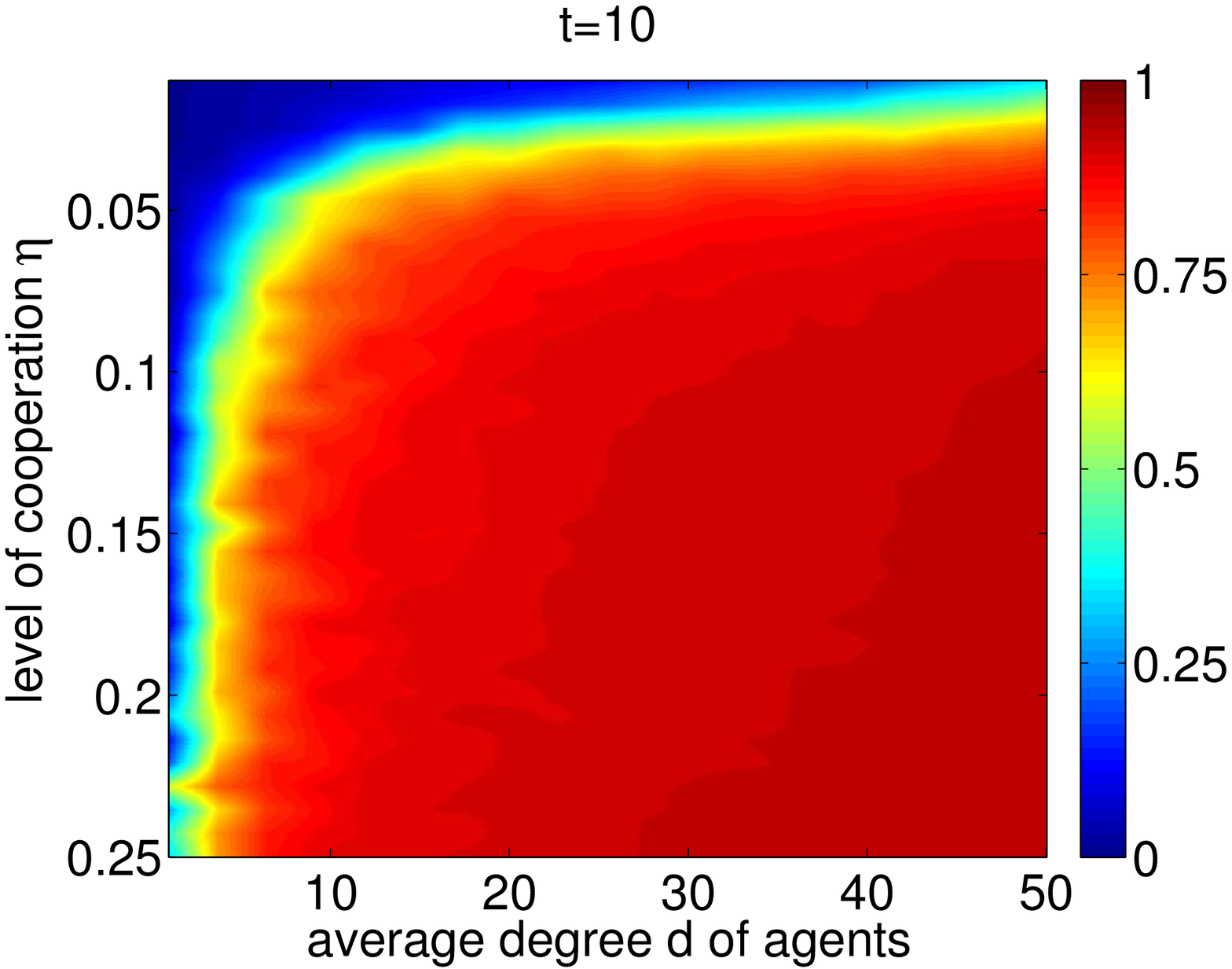}
  \caption{Trust between agents of the same profile as a function of
    level of cooperation and average degree of agents at $t=5$ (left),
    and $t=10$ (right).}
  \label{fig:trust-d-n}
\end{figure}

Figure~\ref{fig:trust-d-n} illustrates the trust between agents of the
same profile as a function of the level of cooperation and the average
degree of agents at $t=5$, and $t=10$. Initially, at $t=0$, agents
still have to learn who to trust (and the whole figure would be blue,
corresponding to zero trust between everyone). At $t=5$, trust is
already developing; for larger average degrees of agents $d$ as well
as for larger levels of cooperation $\eta$, this happens faster. At
$t=10$, trust between agents of the same profile has developed for an
average degree of agents $d>5$ and a level of cooperation $\eta>0.05$.

The obvious consequence of the evolution of trust is that predictions
tend to match the profiles. We test this by measuring the performance
of the system. Let the performance be defined as the sum of the
products of the utility and the trust between all pairs of agents $i$
and $j$:
\begin{eqnarray}
  \label{eq:performance}
  \Phi = \frac{1}{n}\sum_{i}\sum_{j}u_{ij}\frac{T_{ij}}{\sum_{k}T_{ik}},
\end{eqnarray}
where $n$ is the number of agents, e.g.~in our case $n=500$. Agents
are exposed to ratings which lead to both positive or negative
utility. By building trust, they give more weight to the positive
utility and less weight to the negative utility. Therefore, this
measures ``how well agents use their trust''.

\begin{figure}[h]
  \centering
  \includegraphics[width=0.4\textwidth]{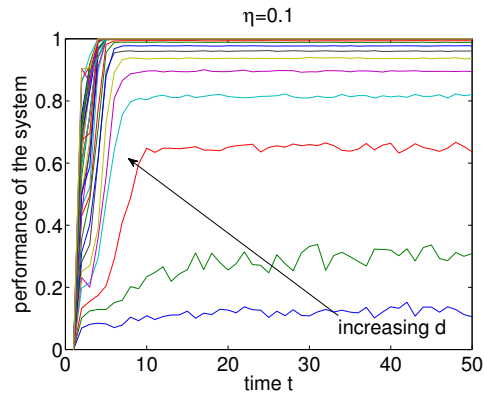}
  \vspace{-1.5ex}
  \caption{Performance over time, for a variable average degree of
    agents, but a fixed level of cooperation.}
  \label{fig:performance-d-n-time}
\end{figure}

Figure~\ref{fig:performance-d-n-time} illustrates the performance over
time: again, the average degree of agents is fixed, $d=7$, and the
level of cooperation $\eta$ is variable, ranging from $0.01$ to $0.25$
in steps of $0.01$. The performance converges to $1$ for almost all
$d$. The similarity to Figure~\ref{fig:trust-d-n-time} is due to the
fact that agents who have developed trust to other agents of the same
profile are provided with good recommendations from their neighbours;
thus, these agents perceive high utility which leads to high
performance.

\begin{figure}[h]
  \centering
  \includegraphics[width=0.4\textwidth]{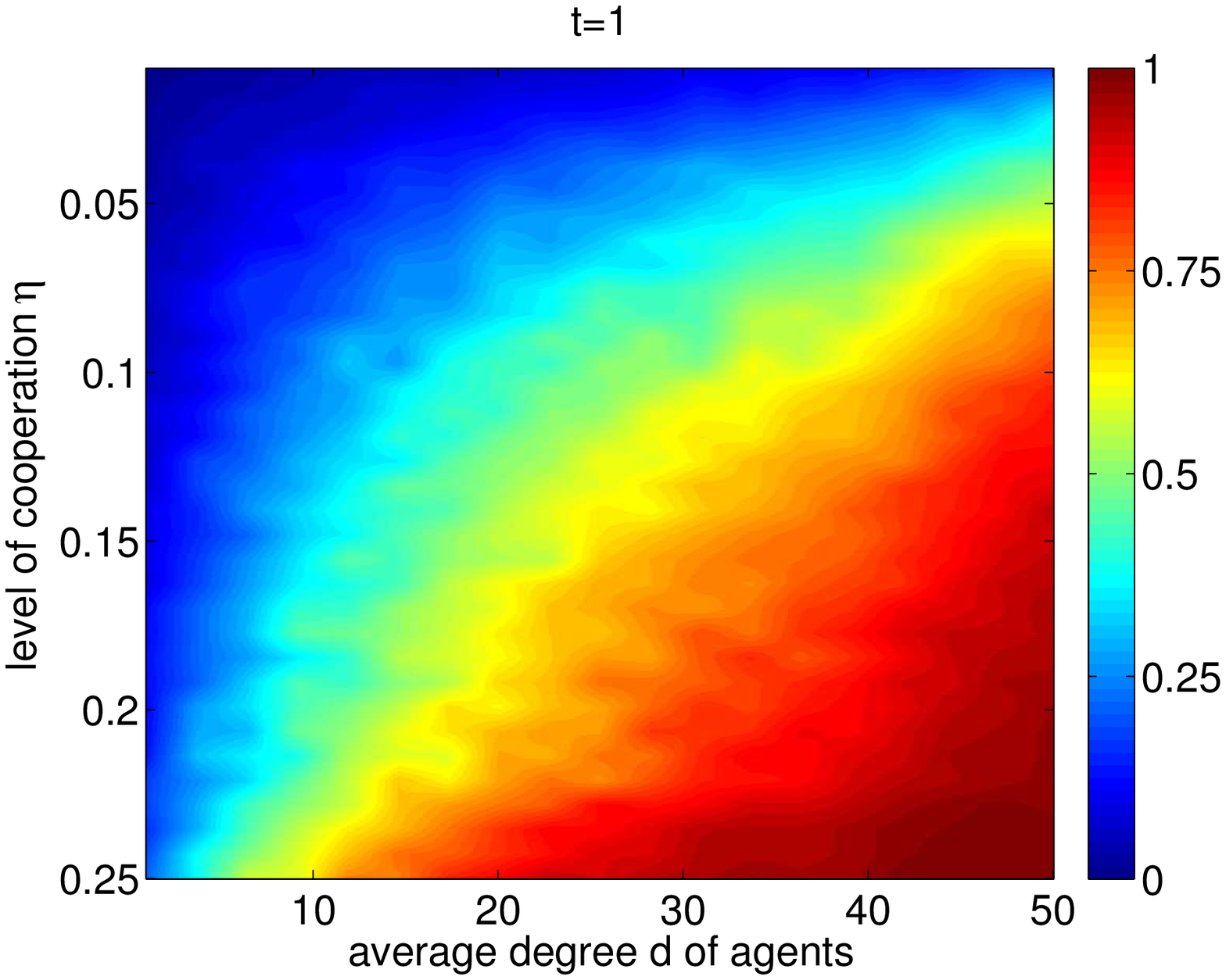}
  \includegraphics[width=0.4\textwidth]{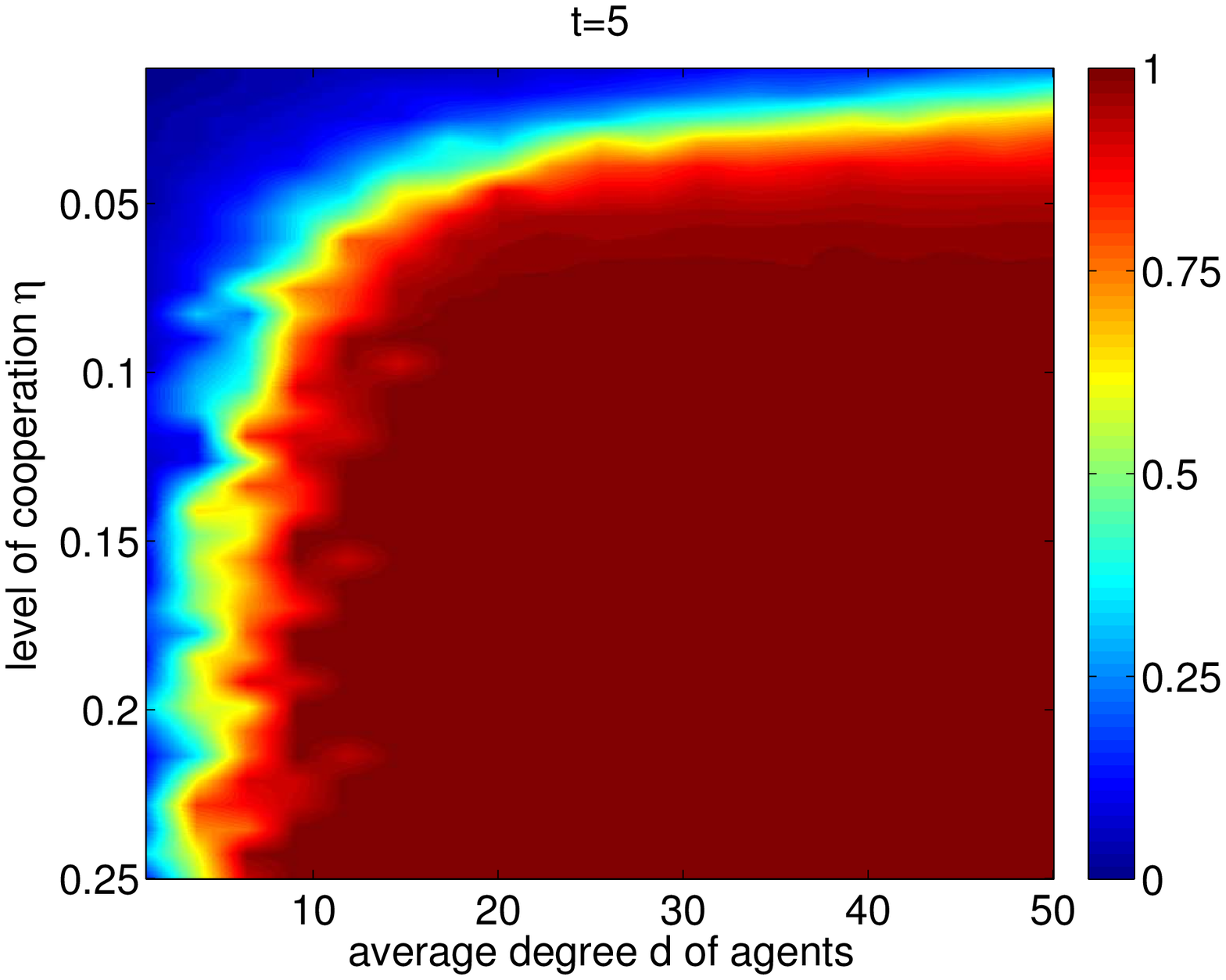}
  \caption{Performance as a function of level of cooperation and
    average degree of agents at $t=1$ (left) and at $t=5$ (right).}
  \label{fig:performance-d-n}
\end{figure}

Finally, Figure~\ref{fig:performance-d-n} illustrates the performance
as a function of the level of cooperation and the average degree of
agents at $t=1$ and at $t=5$. Again, just as the trust between agents
of the same profile increases in Figure~\ref{fig:trust-d-n}, the
performance increases with increasing average degree of agents and
level of cooperation. One might wonder how, at $t=1$, the performance
can already be nonzero -- this is due to the fact that there are only
two opposite profiles; this implies that half of the neighbours of an
agent are of the same profile and, as soon as an agent has developed
some trust to one of these neighbours, it will benefit from their
recommendations which, again, drives the performance up.

\subsection{Empirical Validation}
\label{sec:empirical-validation}

To support the analytical approximations of the model and the results
of the computer simulations, we empirically tested the performance of
a recommender system using our TrustWebRank (TW) metric against one
using a standard Collaborative Filtering (CF) approach, similarly to
what has been done in \cite{massa07}. We crawled Epinions.com, an
on-line platform which allows consumers to read and write reviews
about products. The unique feature of Epinions is that users can also
form a ``web-of-trust'' and specify other users that they trust with
respect to their reviews. The crawling was performed in mid-2007 and
led to a dataset of 60,918 users with 896,969 reviews on 223,687
products and with 518,505 relationships. We cleaned this dataset and
removed users that either had not written any reviews or had no
relationships to other users because no reasonable validation can be
done with these users. Furthermore, we focus on the greatest strongly
connected component (SCC) because a) there is only one large SCC and
many small SCC (1-3 users) and b) membership in this SCC can be seen
as a proxy for having a properly formed web of trust. Having applied
this procedure, we are left with 29,478 users, 731,220 reviews on
201,674 products, and 471,888 relationships. The data sparsity is
99.9877\
stars (max). There is a bias to review favourably, as 75\
ratings are either 4 or 5 stars and only 25\
or 3 stars -- probably because users are more likely to spend time to
write a review when they like a product.

We split the reviews into a training set $R_{\mathrm{Training}}$ and a
test set $R_{\mathrm{Test}}$. We then compare the performance of TW
and CF by training the algorithms on $R_{\mathrm{Training}}$ and
testing with $R_{\mathrm{Test}}$. TW, in general, has comparable
performance to CF, and performs better in particular situations, as we
will describe in the following. The complete empirical validation
will, together with some statistical analyses of the Epinions
community, be reported on in a separate paper
\cite{walter09-epinions}.

\textbf{Mean Absolute Error}: the mean absolute error (MAE) is defined
as

\begin{eqnarray}
  e_{\mathrm{MAE}} = \frac{1}{|R_{\mathrm{Test}}|} \sum_{R_{\mathrm{Test}}}|r_i^o-p_i^o|.
  \label{eq:mae}
\end{eqnarray}

\begin{figure}[tb]
  \centering
  \includegraphics[width=0.4\textwidth]{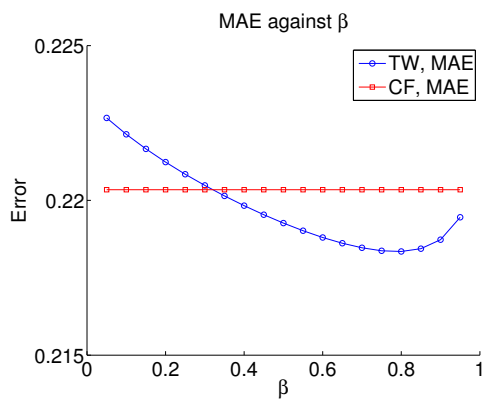}
  \caption{Mean Absolute Error of TW (blue/circles) against $\beta$
    and CF (red/squares). The MAE is normalised to a scale in $[0,1]$,
    i.e.~it reflects percentages.}
  \label{fig:mae}
\end{figure}

Figure \ref{fig:mae} shows the MAE of TW for changing $\beta$ and
CF. Depending on the value of $\beta$, TW performs (marginally) better
than CF. There is an optimal $\beta_{\mathrm{opt}} \approx 0.8$.

However, the fact that most ratings are 4 or 5 limits the meaning of
the MAE as a measure of performance. Indeed, predictions based on the
Simple Average (SA) of ratings on a product, a global algorithm which
is not personalised for users, outperform both TW and CF:
$e_{\mathrm{MAE}}(SA)=0.21$. Similar results were found in
\cite{massa07} using a different dataset of Epinions (from 2003). An
explanation for this is that reviews are very homogeneous and almost
all ratings are positive. Other datasets, such as the commonly used
MovieLens dataset, are more heterogeneous and SA performs worse than
CF on such datasets. Unfortunately, at the moment, Epinions is the
only available dataset which combines rating data and a social network
-- and which is thus suitable to test the performance of TW.

\textbf{Coverage}: coverage measures the percentage of elements that
can be predicted from the training set. Both TW and CF cannot compute
predictions for all elements in the test set. For example, if there is
no similar or trusted user who has rated a particular product, CF or
TW are not able to compute a prediction for that product. CF was able
to compute 41.65\
75.11\
with CF. The reason for this is that TW is able to reach a large
neighbourhood even when the neighbourhood based on co-ratings, as in
CF, is small.

\textbf{Top-N Set Overlap}: as noted, the value of ratings in Epinions
does not seem to carry a lot of meaning -- probably because people
tend to rely more on the text of reviews than on the
rating. Therefore, it makes sense to compare the performance based on
the ability to predict the subset of products rated by a user. We
define the following measures of overlap between sets:

\begin{eqnarray}
  o_i^{N} = \frac{|P_i \cap R^{N}|}{\min(|P_i|,N)} \quad \mathrm{and} \quad o_{X,i}^{N} = \frac{|P_i \cap R_{X,i}^{N}|}{\min(|P_i|,N)} 
  \label{eq:overlap-user}
\end{eqnarray}

where $P_i$ is the set of products rated by a user $i$; $R^{N}$ is the
set of the N most rated products overall in the system; $X$ denotes
either CF or TW and thus $R_{CF,i}^{N}$ and $R_{TW,i}^{N}$ are the
sets of the N most rated products in the neighbourhood of a user $i$
constructed by CF and TW. Note that $R^{N}$ is a global set which is
the same for all users $i$. Thus, $o_i^N$ is the counterpart of
$e_{\mathrm{MAE}}(SA)$ in this context. $R_{CF,i}^{N}$ and
$R_{TW,i}^{N}$ are personalised sets which depend on the neighbourhood
of user $i$ and thus are different for any two users. We define the
average overlap across all users as $O^N$, $O^N_{CF}$, and
$O^{N}_{TW}$. For $N=100$, we obtain $O^{N} \approx 0.0819$,
$O_{CF}^{N} \approx 0.2526$ and $O_{TW}^{N} \approx 0.1724$. Since a
larger overlap signifies a better prediction, the larger the values,
the better the performance. This implies that the global measure
$O^N$ performs worse than both $O_{CF}^N$ and $O_{TW}^N$. In
addition, CF performs better than TW. However, it should be emphasised
that this measure is obviously biased in favour of CF: by definition,
$P_i \cap R^N_{CF,i} \neq \emptyset$. In contrast, $P_i \cap
R^N_{TW,i}$ can be empty, as a user does not necessarily declare trust
to people who are have rated the same items. Still, TW performs
significantly better than the global measure $O^N$.

This illustrates the difficulty to compare the performance of TW with
CF. In fact, the most appropriate way to measure performance would be
based on user-provided feedback subsequent to having followed a
recommendation.

In conclusion, we found that TW and CF have comparable performance. TW
seems mostly useful for recommendations of items different from those
a user has already rated -- e.g.~recommendations on travel books for
people usually interested in tools for gardening.

\section{Extensions and Conclusion}
\label{sec:extensions-conclusion}

We introduced a novel metric for computing indirect trust in social
networks. We derived this metric from feedback centrality measures in
graphs and illustrated how it addresses some limitations of other
trust metrics; most importantly, that it takes cycles in the
underlying graph into account. We constructed a simple model of a
recommender system that makes use of our metric and showed how
indirect trust can be used to generate recommendations. We performed
analytical approximations and computer simulations to characterise the
system behaviour. Finally, we also tested the model by validating it
with empirical data of an Internet community devoted to product
reviews.

Some extensions to this model could involve changing the trust
dynamics:

\textit{Trust update as a slow-positive, fast-negative dynamics.}  It
has been observed in the literature that trust follows a
slow-positive, fast-negative dynamics
\cite{abdul-rahman00,grandison00,marsh94,sabater05,walter08-jaamas}. This
means that trust builds up slowly, but gets torn down quickly and this
behaviour could be implemented by modifying
Eq.~(\ref{eq:trust_update}).

\textit{Coupling the utility with the level of cooperation $\eta$.}
In real applications, if, initially, the utility for users is zero,
then nobody will signal and this is a fixed point -- and a social
dilemma \cite{hardin68}. Thus, we could couple the probability of
signalling to the utility and investigate how to make the system
escape from this undesirable fixed point.

With this work, we have shown that incorporating this novel trust
metric in recommender systems is a promising and viable approach.

\bibliographystyle{acm}
\bibliography{paper}

\begin{thebibliography}{10}

\bibitem{abdul-rahman00}
{\sc Abdul-Rahman, A., and Hailes, S.}
\newblock Supporting trust in virtual communities.
\newblock In {\em Proc. 33th Annual Hawaii Int. Conf. on System Sciences\/}
  (2000), IEEE.

\bibitem{bollobas85}
{\sc Bollob{\'a}s, B.}
\newblock {\em Random Graphs}.
\newblock Academic Press, 1985.

\bibitem{brandes05}
{\sc Brandes, U., and Erlebach, T.}, Eds.
\newblock {\em Network Analysis}.
\newblock Springer, 2005.

\bibitem{brin98}
{\sc Brin, S., and Page, L.}
\newblock The anatomy of a large-scale hypertextual {W}eb search engine.
\newblock {\em Computer Networks and ISDN Systems 30}, 1-7 (1998), 107--117.

\bibitem{cattuto07}
{\sc Cattuto, C., Loreto, V., and Pietronero, L.}
\newblock Collaborative tagging and semiotic dynamics.
\newblock {\em PNAS 104}, 5 (2007), 1461--1464.

\bibitem{erdos59}
{\sc Erd{\H o}s, P., and R{\'e}nyi, A.}
\newblock On random graphs.
\newblock {\em Publ. Math. Debrecen 6\/} (1959), 290--291.

\bibitem{golbeck05}
{\sc Golbeck, J.~A.}
\newblock {\em Computing and applying trust in web-based social networks}.
\newblock PhD thesis, University of Maryland at College Park, 2005.

\bibitem{goldberg92}
{\sc Goldberg, D., Nichols, D., Oki, B.~M., and Terry, D.}
\newblock Using collaborative filtering to weave an information tapestry.
\newblock {\em CACM 35}, 12 (1992), 61--70.

\bibitem{golder05}
{\sc Golder, S., and Huberman, B.~A.}
\newblock The structure of collaborative tagging systems, 2005.

\bibitem{grandison00}
{\sc Grandison, T., and Sloman, M.}
\newblock A survey of trust in internet applications.
\newblock {\em IEEE Communications Surveys and Tutorials 3}, 4 (2000).

\bibitem{hardin68}
{\sc Hardin, G.}
\newblock The tragedy of the commons.
\newblock {\em Science 162\/} (1968), 1243--1248.

\bibitem{herlocker99}
{\sc Herlocker, J.~L., Konstan, J.~A., Borchers, A., and Riedl, J.}
\newblock An algorithmic framework for performing collaborative filtering.
\newblock In {\em SIGIR '99\/} (1999), ACM, pp.~230--237.

\bibitem{horn90}
{\sc Horn, R.~A., and Johnson, C.~R.}
\newblock {\em Matrix Analysis}.
\newblock Cambridge University Press, 1990.

\bibitem{kamvar03}
{\sc Kamvar, S.~D., Schlosser, M.~T., and Garcia-Molina, H.}
\newblock The {Eigentrust} algorithm for reputation management in {P2P}
  networks.
\newblock In {\em WWW '03\/} (2003), ACM, pp.~640--651.

\bibitem{marsh94}
{\sc Marsh, S.}
\newblock {\em Formalising Trust as a Computational Concept}.
\newblock PhD thesis, University of Stirling, 1994.

\bibitem{massa06}
{\sc Massa, P.}
\newblock {\em Trust-aware Decentralized Recommender Systems}.
\newblock PhD thesis, Trento University, 2006.

\bibitem{massa07}
{\sc Massa, P., and Avesani, P.}
\newblock Trust-aware recommender systems.
\newblock In {\em Recommender Systems 2007\/} (2007), ACM.

\bibitem{montaner02b}
{\sc Montaner, M., L{\'o}pez, B., and {de la Rosa}, J.~L.}
\newblock Opinion-based filtering through trust.
\newblock In {\em Proc. 6th Int. Workshop on Cooperative Information Agents\/}
  (2002), Springer, pp.~164--178.

\bibitem{montaner03a}
{\sc Montaner, M., L{\'o}pez, B., and {de la Rosa}, J.~L.}
\newblock A taxonomy of recommender agents on the internet.
\newblock {\em Artificial Intelligence Review 19}, 4 (2003), 285--330.

\bibitem{newman02}
{\sc Newman, M. E.~J., Watts, D.~J., and Strogatz, S.~H.}
\newblock Random graph models of social networks.
\newblock {\em PNAS 99}, 90001 (2002), 2566--2572.

\bibitem{sabater05}
{\sc Sabater, J., and Sierra, C.}
\newblock Review on computational trust and reputation models.
\newblock {\em Artificial Intelligence Review 24}, 1 (2005), 33--60.

\bibitem{sarwar01}
{\sc Sarwar, B., Karypis, G., Konstan, J., and Riedl, J.}
\newblock Item-based collaborative filtering recommendation algorithms.
\newblock In {\em WWW '01\/} (2001), ACM, pp.~285--295.

\bibitem{seneta06}
{\sc Seneta, E.}
\newblock {\em Non-Negative Matrices and Markov Chains}.
\newblock Springer, 2006.

\bibitem{vega-redondo07}
{\sc Vega-Redondo, F.}
\newblock {\em Complex Social Networks}.
\newblock Cambridge University Press, 2007.

\bibitem{walter08-jaamas}
{\sc Walter, F.~E., Battiston, S., and Schweitzer, F.}
\newblock A model of a trust-based recommendation system on a social network.
\newblock {\em Journal of Autonomous Agents and Multi-Agent Systems 16}, 1
  (2008), 57--74.

\bibitem{walter09-epinions}
{\sc Walter, F.~E., Battiston, S., and Schweitzer, F.}
\newblock An investigation of {Epinions.com}: Statistical properties and
  empirical validation of recommender system algorithms, 2009.
\newblock In progress.

\bibitem{wassermann94}
{\sc Wassermann, S., and Faust, K.}
\newblock {\em Social Network Analysis: Methods and Applications}.
\newblock Cambridge University Press, 1994.

\bibitem{ziegler06}
{\sc Ziegler, C.-N., and Golbeck, J.}
\newblock Investigating correlations of trust and interest similarity.
\newblock {\em Decision Support Systems\/} (2006).

\end{thebibliography}

\end{document}